\documentclass[pdflatex,sn-mathphys-num,iicol]{sn-jnl}
\usepackage{graphicx}
\usepackage{amsmath,amssymb,amsfonts}
\usepackage{booktabs}
\usepackage{xcolor}
\usepackage{textcomp}
\usepackage{bookmark}
\usepackage{xurl}
\hypersetup{hypertexnames=false}
\begin{document}

\title[No own-model premium in AI--AI bias]{A Shared Taste for Model-Written Text:
The Generator-by-Selector Matrices of ``AI--AI Bias'' Show No Detectable
Own-Model Premium}

\author*[1,2]{\fnm{Dmitrij} \sur{\.Zatuchin}}\email{dmitrij.zatuchin@eek.ee}

\affil*[1]{\orgdiv{Department of Information Technologies}, \orgname{Estonian
Entrepreneurship University of Applied Sciences (EUAS)},
\orgaddress{\city{Tallinn}, \country{Estonia}}}
\affil[2]{\orgname{Rankfor.AI}, \orgaddress{\city{Wroc\l{}aw}, \country{Poland};
\city{Tallinn}, \country{Estonia}}}

\abstract{Laurito et al.\ (PNAS 2025) showed that large language models
choosing between two descriptions of the same product, paper or film prefer
the description written by a language model over the one written by a person,
by a wide margin over what human judges do. Their design crosses five
generators with the same five models as selectors, which permits a second
question the paper does not headline: does a selector prefer text from its own
model beyond what the generator and selector main effects predict? We rebuild
the three 5$\times$5 matrices from the per-item counts in the authors' public
repository (21,828 valid trials; every cell matches the published value) and
fit a two-way fixed-effects model with an own-model term $\gamma$,
tested by the exact permutation test over the 120 relabellings of the
selectors. The premium is $+0.013$ on products (exact one-sided $p = 0.24$),
$-0.010$ on paper abstracts ($p = 0.74$), $+0.054$ on films ($p = 0.07$) and
$+0.019$ pooled ($p = 0.14$; 95\% interval $-0.008$ to $0.046$). The
same-vendor term for the GPT-3.5 and GPT-4 pair is negative in all three
datasets. Position bias moves single cells by up to 0.42 share points in either direction, and the own-model contrast is unchanged
once order-driven items are removed. The design would have detected a premium
of 0.05 with 82\% (products), 88\% (papers), 42\% (films) and 97\% (pooled)
power; the minimum detectable effect at 80\% power is 0.034 pooled. The
absence is informative down to about 0.04 share points and silent below that.
The 4$\times$4 matrix of Tan et al.\ (ACL 2024) gives $\gamma =
+0.148$ at the smallest $p$ its 24 relabellings allow, with a same-family term
of the same size. The main result of Laurito et al.\ stands: models share a
taste for model-written text, with GPT-4's descriptions chosen 77\% to 95\% of
the time by every selector on products. What these data do not show is a
model recognising and favouring its own prose.}


\maketitle

\section{Introduction}\label{sec:intro}

Laurito et al.\ asked whether large language models, used as decision
assistants, favour goods pitched in language-model prose over the same goods
pitched by a person \cite{laurito2025aiaibias}. For 109 products, 100 paper
abstracts and 250 film plot summaries, each item was described once by a human
and once by each of five models, and each of the same five models then chose
between the two descriptions in a binary prompt shown in both orders. Every
selector chose the model-written product description more often than the human
judges did, by 26 to 59 share points over the ten cells with a human
comparison. The finding, which the authors call AI--AI bias, has since drawn a published letter and reply
\cite{yu2025broader,laurito2025reply}.

The design also answers a question the paper leaves in its figures. Because the
generators and the selectors are the same five models, the results form three
5$\times$5 matrices, and a matrix can be decomposed: how much of a cell is the
generator's text being attractive to every selector, how much is the selector
leaning toward model text in general, and how much is left on the diagonal,
where a model judges its own output. That diagonal is the quantity the
self-preference literature is about \cite{panickssery2024recognize,
wataoka2024selfpreference}, and it matters in practice. If a selector favours
its own model's prose, then a text tuned for one engine is tuned against the
others; if the preference is shared, one model-friendly text serves them all.

This note rebuilds the matrices from the authors' repository and tests the
diagonal. It adds nothing to the original finding and takes nothing from it.
The contributions are: (i) the three matrices rebuilt from per-item counts, with
every cell matching the published value; (ii) the own-model premium per dataset
and pooled, with an exact permutation test, a standard error, and the
same-vendor effect separated from it; (iii) the effect of position bias on the
diagonal, from the runs in which the repository records which position the
selector chose; (iv) the minimum detectable effect of this design and a power
table for a larger one; and (v) the same test on the one comparable published
matrix, Tan et al.\ \cite{tan2024blinded}, and the placement of both results in
the 2024 to 2026 literature on self-preference.

\section{Data}\label{sec:data}

\textbf{Source and licence.} The article is published under CC BY 4.0 and
states that all study data are included in the article and its SI Appendix.
The code repository named in the article \cite{laurito2025repo} carries no
licence file; under GitHub's terms the code is therefore readable and
forkable and no further right is granted. We ran no code from it. The
repository stores, for every run, one JSON record per item with the number of
orderings in which the selector chose the model text, the human text, or gave no
valid choice, and, for the runs of January 18 to 30, 2025, the number of times
the first-listed text was chosen. No per-trial rows with position exist for the
other runs. All quantities in this note are therefore aggregates over items or
over cells, and the derived tables we release contain no text from either
source.

\textbf{Design as recorded.} For each dataset the paper uses one generation
prompt and one selection prompt; the repository names them, and 2025 runs of
that pair exist for all 75 cells. The paper names GPT-3.5-turbo-0125 for
products and films and GPT-3.5-turbo-1106 for papers, for both roles; the runs
follow the same split. In the final design each item has one model-written
description shown against the human one in both orders, so a cell holds two
trials per item: 218 or 220 on products, 186 on papers, 500 on films, less the
invalid trials. Papers are the exception in one cell: the GPT-4 text judged by GPT-3.5
was last run in January 2024 with four descriptions per item (744 trials, 124
invalid), and that is the run the published value 0.69 comes from; the January
2025 rerun of the same cell gives 0.71.

\textbf{Cell selection.} For each cell we took the latest run whose LLM-win
share, wins over wins plus losses, rounds to the published value. All 75 cells
matched; the largest deviation between a rebuilt share and its published
rounding is 0.005. The 75 runs hold 21,828 valid trials and 1,280 invalid ones.
Where the repository holds several distinct 2025 runs of a cell (29 cells, 122
distinct runs in all), the others enter a sensitivity analysis in
Section~\ref{sec:replicates}. The paper's dataset of 100 papers appears as 93
items in every paper run. Human selectors judged only the GPT-3.5 and GPT-4
texts; we take those six values from the article's Figures 2 to 4.

\section{Method}\label{sec:method}

\textbf{Model.} Let $y_{ij}$ be the share of valid trials in which selector $j$
chose the text written by generator $i$ over the human text. We fit
\begin{equation}
y_{ij} = \mu + a_i + b_j + \gamma\,[i = j] + e_{ij},
\end{equation}
with generator effects $a_i$, selector effects $b_j$ and the own-model premium
$\gamma$ on the diagonal. In a full $k \times k$ matrix with one self cell per
row and column, the least-squares $\gamma$ equals the mean of the $k$ diagonal
cells minus the mean of the $k(k-1)$ others, so it has the same units as the
cells: share points of the binary choice.

\textbf{Inference.} The null hypothesis is that model identity carries no
information beyond the row and column effects, so the labelling of which
selector counts as ``self'' for each generator is exchangeable. We recompute
$\gamma$ under every relabelling of the five selector columns, 120 in all, and
report the one-sided exact $p$, the share of relabellings with a premium at
least as large as the observed one, together with the two-sided $p$. The
smallest one-sided $p$ this design can return is $1/120 = 0.008$. The pooled
test double-centres each matrix and fits one $\gamma$ to the three stacked
residual matrices under a common relabelling. We also report the least-squares
standard error, $\sigma/\sqrt{k-1}$ per matrix on 15 residual degrees of
freedom and $\sigma/\sqrt{3(k-1)}$ pooled on 45, where $\sigma$ is the residual
standard deviation after all terms. The permutation test makes no
distributional assumption; the standard error gives an interval.

\textbf{Same-vendor effect.} GPT-3.5 and GPT-4 are the only two models from one
vendor, so the two cells where one judges the other's text carry a family
dummy, fitted with the self dummy and permuted with it.

\textbf{Trial-level model.} To show what a trial-level analysis would claim, we
fit a binomial logistic regression on the per-item counts with item, generator
and selector fixed effects and the two dummies, with standard errors clustered
by item, and we compute the cell-level Pearson dispersion of that fit: the
ratio of the observed cell variance to what binomial sampling of the fitted
shares predicts.

\textbf{Position.} For the 27 runs that record first-listed choices, the two
trials of an item with no invalid response fall into four patterns: the model
text chosen in both orders, the human text chosen in both, the first-listed
text chosen in both, or the second-listed in both. The last two are decided by
position, the first two by content. We report each run's first-choice rate, the
share of items decided by position, and the LLM-win share among content-decided
items, and we compare the diagonal to the off-diagonal cells of the same row on
both measures. Items with more than two trials whose position split is not
determined by the recorded counts (62 of 3,645) are left out of this analysis.

\textbf{Power.} Residual variance in a matrix has two parts: binomial sampling
of the cell share, $p_{ij}(1-p_{ij})/n_{ij}$, which more trials shrink, and
cell-level interaction noise, which they do not. We estimate the second as the
residual variance of the fitted model less the mean binomial variance. The
minimum detectable effect is then found by simulation: to the additive fit of
each matrix we add a premium $\gamma$ on the diagonal, Gaussian interaction
noise with the estimated standard deviation, and binomial sampling at the
observed trial counts, and we run the exact test on each draw. The reported
value is the $\gamma$ at which the one-sided test at $\alpha = 0.05$ rejects in
80\% of draws. The same machinery, with the pooled interaction estimate, a base
share of 0.7 and random row and column effects, gives the power table for
larger designs; for $k > 6$ the null uses 600 random relabellings. All
simulation is seeded; the analysis script and the extracted counts are in the
package.

\section{Results}\label{sec:results}

\subsection{The own-model premium}\label{sec:gamma}

\begin{table*}[ht]
\caption{Own-model premium in the three matrices of Laurito et al.\ and in Tan
et al. $\gamma$ is the mean of the diagonal cells minus the mean of the others
after row and column effects; $p$ is the exact one-sided permutation $p$ over
all relabellings of the selectors (120 for $k=5$, 24 for $k=4$); the interval is
$\pm t_{0.975}$ times the least-squares standard error. ``Self is row max''
counts generators whose own selector gave their text its highest share. Family
is the same-vendor term fitted alongside $\gamma$ (GPT-3.5 and GPT-4; Llama 2 7B
and 13B for Tan et al.).}\label{tab:gamma}
\centering
\footnotesize
\setlength{\tabcolsep}{4pt}
\begin{tabular}{lrrrrrlrr}
\toprule
Matrix & Trials & Diag. & Off-diag. & $\gamma$ & Exact $p$ & 95\% interval & Row max & Family ($p$) \\
\midrule
Products, $5\times5$ & 5,419 & 0.788 & 0.775 & $+0.013$ & 0.242 & $-0.026$ to $0.053$ & 1 of 5 & $-0.012$ (0.75) \\
Paper abstracts, $5\times5$ & 4,779 & 0.605 & 0.615 & $-0.010$ & 0.742 & $-0.046$ to $0.027$ & 1 of 5 & $-0.016$ (0.68) \\
Movie summaries, $5\times5$ & 11,630 & 0.674 & 0.620 & $+0.054$ & 0.067 & $-0.013$ to $0.121$ & 2 of 5 & $-0.039$ (0.71) \\
Pooled, three matrices & 21,828 & & & $+0.019$ & 0.142 & $-0.008$ to $0.046$ & 4 of 15 & $-0.022$ (0.72) \\
\midrule
Tan et al.\ 2024, $4\times4$ & & 0.595 & 0.447 & $+0.148$ & 0.042 & $+0.001$ to $+0.296$ & 3 of 4 & $+0.163$ (0.17) \\
\bottomrule
\end{tabular}
\end{table*}

\begin{figure*}[ht]
\centering
\includegraphics[width=\textwidth]{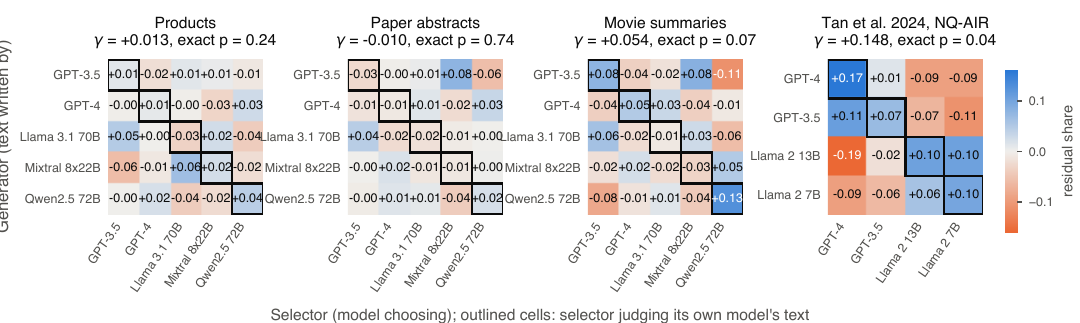}
\caption{Residual share after removing generator and selector effects, for the
three matrices of Laurito et al.\ and the NQ-AIR matrix of Tan et al.\ (bias
toward the generated over the retrieved context). Outlined cells are the
diagonal, where the selector judges its own model's text. Blue is above the
additive prediction, orange below.}\label{fig:residuals}
\end{figure*}

Table~\ref{tab:gamma} gives the premium, and Figure~\ref{fig:residuals} shows
the residual matrices with the diagonal outlined. On products the five
diagonal cells average 0.788 against 0.775 for the twenty others, a premium of
$+0.013$; 29 of the 120 relabellings do at least as well, $p = 0.24$. On paper
abstracts the diagonal is below the rest, $\gamma = -0.010$, $p = 0.74$. On
films the premium is $+0.054$ with $p = 0.07$: eight relabellings of 120 reach
it, and the 95\% interval runs from $-0.013$ to $0.121$. Pooled over the three
matrices $\gamma = +0.019$, $p = 0.14$, interval $-0.008$ to $0.046$. In only 4
of the 15 generator rows is the generator's own model the selector that likes
its text most. The same test on the figure values as printed to two decimals gave
$+0.0125$, $-0.010$, $+0.050$ and $+0.018$ with $p$ of 0.27, 0.74, 0.08 and
0.18; the rebuilt counts move nothing of substance.
On the logit scale, which is the natural scale if position bias compresses
shares multiplicatively toward 0.5 (Section~\ref{sec:position}), the pooled
premium is $+0.049$ log-odds with $p = 0.27$.

\begin{figure*}[ht]
\centering
\includegraphics[width=\textwidth]{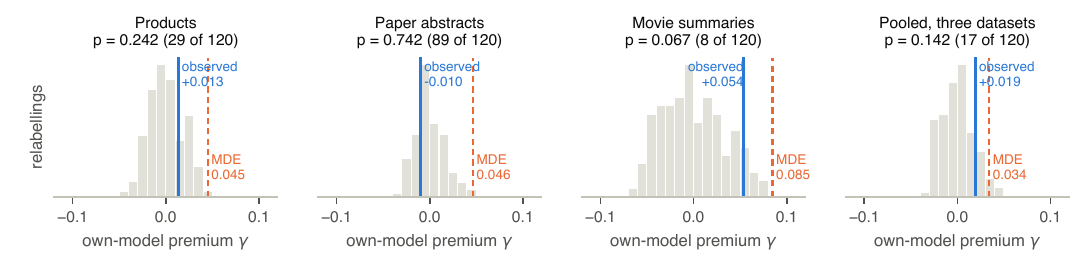}
\caption{Exact permutation null of the own-model premium: $\gamma$ under each
of the 120 relabellings of the selectors, with the observed value (solid) and
the minimum detectable effect at 80\% power (dashed).}\label{fig:perm}
\end{figure*}

\subsection{Same vendor}\label{sec:family}

The one pair of models from a single vendor gives two off-diagonal cells per
matrix: GPT-3.5 judging GPT-4's text and the reverse. Fitted alongside the self
term, the family coefficient is $-0.012$ on products, $-0.016$ on papers and
$-0.039$ on films, $-0.022$ pooled ($p = 0.72$), so those two cells sit below
what the row and column effects predict, in all three datasets. With one pair the test has
no power to speak of, and the design cannot separate a vendor effect from a
size effect within a vendor; the direction does not favour a same-vendor
premium.

\subsection{Trial-level inference and why it overstates}\label{sec:replicates}

A binomial regression on the 11,300 item-by-cell counts, with item, generator
and selector fixed effects and standard errors clustered by item, returns a
self coefficient of $-0.035$ log-odds on products (SE 0.079), $-0.077$ on
papers (0.064) and $+0.256$ on films (0.045), $+0.135$ pooled (0.034, $z =
3.9$). At face value the film and pooled estimates are highly significant, and face
value is the wrong reading. The Pearson dispersion of
the fitted cell shares is 3.7 on products, 1.1 on papers, 7.4 on films and 4.0
pooled: cells vary three to seven times more than binomial sampling of their
trials predicts, so the trial is the wrong unit and the cell is the right one.
The residual standard deviation per cell is 0.037 on products, 0.034 on papers
and 0.063 on films; the part binomial sampling explains is 0.027, 0.036 and
0.022, leaving cell-level noise of 0.026, nil, and 0.059. Films are the noisy
dataset, and papers are the one where more trials would help.

The repository shows where the film noise comes from. For 29 cells it holds
more than one distinct 2025 run of the same prompt pair; the median spread
between runs of one cell is 0.012 share points, and the largest is 0.161, for
GPT-4's film summaries judged by GPT-3.5. That selector returned 22 invalid
choices in one run of that cell, 94 in a second and 142 in the published one,
and the share moved from 0.54 to 0.66 to 0.70. Recomputing $\gamma$ with the
mean over distinct runs per cell gives $+0.011$, $-0.010$ and $+0.054$ ($p =
0.30$, 0.72 and 0.08); with the latest run per cell, $+0.005$, $-0.010$ and
$+0.052$ ($p = 0.40$, 0.69 and 0.09). The conclusion does not depend on which
run stands for a cell.

\subsection{Position bias}\label{sec:position}

\begin{table}[ht]
\caption{Position and the diagonal, in the rows where the repository records
first-listed choices. ``Raw'' is the LLM-win share over all valid trials; ``content''
is the share among items decided the same way in both orders; ``by position''
is the share of items decided by order. Contrast is the own-model cell minus
the mean of the other cells in the row.}\label{tab:position}
\centering
\footnotesize
\setlength{\tabcolsep}{1.5pt}
\begin{tabular}{lrrrr}
\toprule
& \multicolumn{2}{c}{Contrast, self $-$ others} & \multicolumn{2}{c}{By position} \\
Row & raw & content & self & others \\
\midrule
Products, Llama 3.1 70B & $-0.195$ & $+0.019$ & 0.72 & 0.26 \\
Products, Mixtral 8x22B & $+0.121$ & $+0.002$ & 0.26 & 0.58 \\
Papers, GPT-4 & $-0.024$ & $-0.035$ & 0.28 & 0.26 \\
Films, GPT-4 & $-0.006$ & $-0.088$ & 0.46 & 0.53 \\
\midrule
Mean & $-0.026$ & $-0.025$ & 0.42 & 0.41 \\
\bottomrule
\end{tabular}
\end{table}

\begin{figure}[ht]
\centering
\includegraphics[width=0.95\columnwidth]{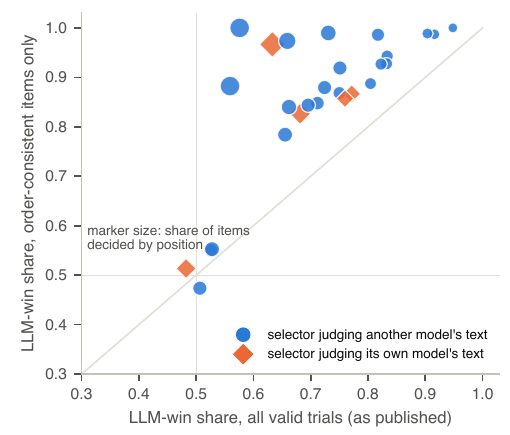}
\caption{LLM-win share with and without the items decided by position, for the
27 runs that record first-listed choices. Points on the diagonal line are
unaffected by position; the vertical distance is the compression toward 0.5
that order habits impose. Own-model cells (diamonds) sit among the
others.}\label{fig:position}
\end{figure}

The authors report first-item rates per selector and note that a strong
position habit can mask a preference, since a selector that always takes the
first text scores 0.5 on any cell. The recorded runs bear this out and add that
the habit runs both ways. Across 27 runs the first-choice rate spans 0.109
(GPT-3.5 judging GPT-4's film summaries, which took the second text nine times
in ten) to 0.923 (Llama 3.1 70B judging Mixtral's product text). The share of
items decided by position runs from 0.08 to 0.84 with a median of 0.40, and in
18 of the 27 runs the model text is chosen at a different rate when listed
first than when listed second (Fisher exact $p < 0.05$). Removing the
position-decided items changes a cell's share by 0.145 on average and by 0.424
at most (Figure~\ref{fig:position}). The GPT-3.5 film selector's two runs
of one cell on one day had first-choice rates of 0.11 and 0.49, the
instability behind Section~\ref{sec:replicates}.

So position bias can move any single cell, including a diagonal one, by a lot.
Whether it can manufacture or hide a diagonal effect depends on whether it
differs on the diagonal, and Table~\ref{tab:position} says it does not. Llama
3.1 70B decides 72\% of its own product descriptions by position against 26\%
of the others, which pulls its own cell down from 0.83 to 0.63 raw, a contrast
of $-0.195$; on content-decided items the contrast is $+0.019$. Mixtral shows
the mirror image, $+0.121$ raw and $+0.002$ on content. GPT-4 on papers and on
films is close to zero either way. Over the four rows the mean contrast is
$-0.026$ raw and $-0.025$ on content, and the share of position-decided items
is 0.42 on the diagonal against 0.41 off it. Position bias is large, selector
specific, and in these rows blind to whose text is being judged. Since every
pair was shown in both orders, its net effect on a cell is compression toward
0.5, and the logit-scale result above is the corresponding check on the full
matrices.

\subsection{What the design could have seen}\label{sec:power}

\begin{figure*}[ht]
\centering
\includegraphics[width=\textwidth]{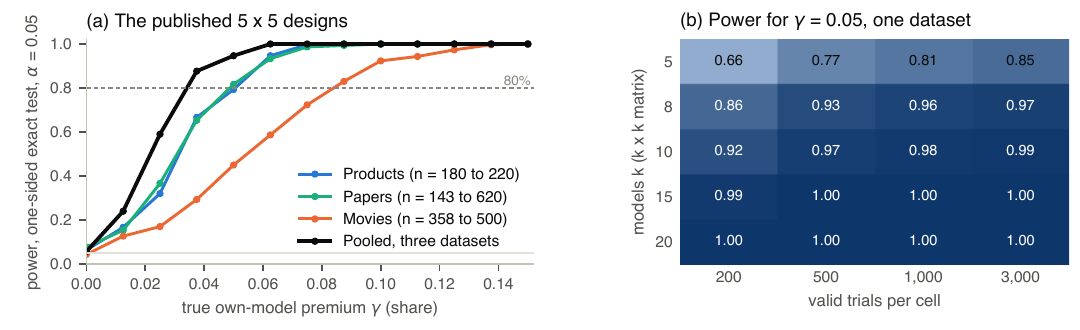}
\caption{(a) Power of the one-sided exact test at $\alpha = 0.05$ against the
true own-model premium, simulated on each matrix's additive fit with its
estimated cell-level noise and its trial counts. (b) Analytic power for a
premium of 0.05 in a $k \times k$ design with $n$ valid trials per cell, one
dataset, cell-level noise 0.037 as estimated here.}\label{fig:power}
\end{figure*}

Figure~\ref{fig:power}a gives the power of the exact test against the true
premium. At $\gamma = 0.05$ the products matrix rejects in 82\% of draws, papers
in 88\%, films in 42\% and the pooled test in 97\%; at $\gamma = 0.10$ all four
reach at least 89\%. The minimum detectable effect at 80\% power is 0.045 on products,
0.046 on papers, 0.085 on films and 0.034 pooled. The one-sided 95\% upper
bound on the pooled premium is 0.042. Read together with the discrete floor of
$p = 1/120$: this design was well placed to see an own-model premium of five
share points, and would almost surely have seen ten; it says nothing about
three or less. For scale, the gap between LLM and human selectors on GPT-4's
product text is 41 to 59 share points (0.77 to 0.95 against 0.36), and the
generator effects span 0.18 to 0.26 share points within a matrix. An
own-model effect, if one exists in this setting, is an order of magnitude
smaller than the shared effect.

\begin{table}[ht]
\caption{Power of the one-sided exact test at $\alpha = 0.05$ for a future
$k \times k$ study, one dataset, cell-level noise 0.037 and base share 0.7.
Analytic value; the simulated value agrees within 0.05 in every cell shown.}\label{tab:power}
\centering
\footnotesize
\begin{tabular}{rrrrr}
\toprule
& \multicolumn{2}{c}{$\gamma = 0.05$} & \multicolumn{2}{c}{$\gamma = 0.10$} \\
$k$ & $n = 200$ & $n = 1{,}000$ & $n = 200$ & $n = 1{,}000$ \\
\midrule
5 & 0.66 & 0.81 & 0.99 & 1.00 \\
8 & 0.86 & 0.96 & 1.00 & 1.00 \\
10 & 0.92 & 0.98 & 1.00 & 1.00 \\
15 & 0.99 & 1.00 & 1.00 & 1.00 \\
20 & 1.00 & 1.00 & 1.00 & 1.00 \\
\bottomrule
\end{tabular}
\end{table}

Table~\ref{tab:power} and Figure~\ref{fig:power}b give the same calculation
for a future study. Because the cell-level noise does not shrink with trials,
adding trials to a 5$\times$5 design buys little: 3,000 trials per cell reach
85\% power for $\gamma = 0.05$ where 200 reach 66\%. Adding models buys more:
eight models at 200 trials per cell give 86\%, ten give 92\%. A premium of 0.10
is detectable with three models and any reasonable trial count. The standard
error of $\gamma$ is $\sqrt{(\sigma_{\mathrm{int}}^2 + p(1-p)/n)/(k-1)}$, and the
first term dominates once $n$ passes a few hundred. A study designed to settle
the own-model question in this paradigm should therefore add generators and
selectors, ideally several per vendor so that the vendor and self terms
separate, and should record position per trial so the compression can be
modelled, which beats removing it.

\subsection{Tan et al.\ beside it}\label{sec:tan}

The one other published matrix of this shape that we know of is Figure 6 of
Tan et al.\ \cite{tan2024blinded}: four readers (GPT-4, GPT-3.5, Llama 2 13B
and 7B) each given a question with two contexts, one retrieved and one written
by each of the four models as generator, and the cell is the bias toward the
generated context on NQ-AIR. Here $\gamma = +0.148$ with $p = 0.042$, which is
$1/24$, the smallest value 24 relabellings allow, and the self cell is the row
maximum for three generators of four. The authors report this as models
preferring self-generated contexts, with Llama 2 7B the exception because it
prefers 13B's text. Two readings temper the number. First, fitted with a family
term, the self and same-family coefficients are $+0.203$ and $+0.163$: the
same-family off-diagonal cells (mean 0.555) are elevated almost as much as the
diagonal over the cross-family cells (0.393), which is a family effect as much
as a self effect, on two families. Second, a reader's own generated context
restates the reader's parametric belief, so ``own text'' and ``agrees with what
I already think'' are one variable there, whereas in Laurito et al.\ the model
text and the human text describe the same item from the same facts. The two
matrices measure different things and do not conflict.

\section{Related work}\label{sec:related}

Self-preference in judging has been measured many ways since 2024, and the
picture is one of a real but small, heterogeneous and largely style-carried
effect. Panickssery et al.\ tied self-preference to self-recognition in
summarisation with three models \cite{panickssery2024recognize}; Wataoka et
al.\ showed that judges score low-perplexity text higher whoever wrote it, which
makes own-output preference a special case of familiarity
\cite{wataoka2024selfpreference}; Davidson et al.\ and Bai et al.\ found no
consistent self-recognition across ten models
\cite{davidson2024selfrecognition,bai2025knowthyself}. Quality-controlled
estimates are smaller than raw ones: Chen et al.\ measure self-preference
against gold judgments and find that rewriting both answers into one style cuts
it by a third to a half \cite{chen2025beyondsurface}; Chen, Wei et al.\ find
that most of the self-preference of strong judges on verifiable tasks is
warranted, because their answers are right \cite{chen2025forareason}; Roytburg
et al.\ show that only about half of previously reported self-preference cases
survive a null built from the judge's own voting distribution
\cite{roytburg2026narcissists}; Chae et al.\ find that it vanishes on three of
four rubric dimensions once selection quality and judge severity are
controlled, while a bare ``self'' label shifts scores on its own
\cite{chae2026bidirectional}; and Yang et al., across 20 judges on
equal-quality pairs, find eight with a positive and nine with a negative
self-preference \cite{yang2026quantifying}. Guey and Bougault, with a
deterministic checker deciding validity, find no self-preference in four
mid-tier families, with effects under 13 points not excluded
\cite{guey2026weakorabsent}. Spiliopoulou et al.\ and Xu et al.\ report
model-specific and heterogeneous self-bias, strongest for a few frontier judges
\cite{spiliopoulou2025playfavorites,xu2025hiring}.

Family effects with the self cells removed are the closest analogue to
Section~\ref{sec:family}. Awuni et al.\ cross four open-weight families with the
judge never scoring its own output and report a same-family lift of 3.4 to 8.4
points after holding candidate family fixed \cite{awuni2026whojudges}; Philipp
et al.\ run nine judges over five generators on clinical items with physicians
as anchor and find sibling lifts of 6.6 and 11.5 points inside the GPT and
Gemma lineages \cite{philipp2026clinician}; Li et al.\ define inheritance and
same-family relatedness and find judges favouring related models
\cite{li2026preferenceleakage}; Pombal et al.\ carry a family variant of their
rubric metric \cite{pombal2026rubric}. Those designs measure positive family
effects of 3 to 12 points in their settings; the Laurito matrices, with one
vendor pair and a negative coefficient, cannot confirm or deny an effect of
that size.

On the retrieval side, where the question is whether a reader prefers
model-written passages and its own in particular, Chen et al.\ find
self-preference under style-based prompts and none under fact-centric RAG
prompts \cite{chen2025biasedevaluators}, and Ross et al.\ find that answer
retention decides which representation a reader benefits from, with no
family alignment \cite{ross2026representations}. Wu et al.\ learn per-engine
preference rules for web content and find 79\% to 84\% of the rules shared
across Gemini, GPT and Claude, with an engine-specific increment of 1\% to 8\%
\cite{wu2026autogeo}. The present note fits that pattern: a large shared effect
and no detectable model-specific one.

\section{Limitations}\label{sec:limits}

The re-analysis inherits the design. Five models, one pair from a single
vendor, one prompt per role, one model text per item, and binary choices with
some selectors returning many invalid answers: GPT-3.5 on films gave no valid
choice in 108 to 142 trials of 500 in each of its five cells. The test is on cell
shares, and the noise model behind the power figures treats cell-level
interaction as Gaussian with a standard deviation estimated on 15 degrees of
freedom per matrix; those power numbers are indicative to perhaps 10 points
either way. Position is recorded for 27 of the 75 cells' worth of runs, and the
diagonal comparison in Table~\ref{tab:position} rests on four rows. The
same-vendor test has two cells per matrix. Nothing here speaks to whether a
model can recognise its own text when asked, to judges scoring open-ended
outputs, or to models released after 2024. The five models are GPT-3.5-turbo,
GPT-4-1106-preview, Llama 3.1 70B, Mixtral 8x22B and Qwen2.5 72B, and the
finding is about them.

\section{Conclusion}\label{sec:conclusion}

Laurito et al.\ established that language models prefer language-model prose
when a person would not, and that stands. Their matrices add one more fact.
Decomposed into what a generator's text does for every selector and what a
selector does for every text, the residue on the diagonal is $+0.019$ share
points pooled, with 17 of 120 relabellings doing as well, an upper bound near
0.04, and no dataset in which the generator's own model is reliably its keenest
selector. The same-vendor pair sits below its additive prediction in all three
datasets, position bias moves cells by up to 0.42 and leaves the diagonal
contrast where it was, and the design had 97\% power to see a premium of 0.05. The
preference these models show is a shared taste: GPT-4's product descriptions
are chosen 77\% to 95\% of the time by all five selectors, GPT-3.5's 55\% to
76\%, and a person chooses them 36\% and 29\% of the time. A model preferring
its own voice would be a different and narrower phenomenon, and one these data
do not contain.

\textbf{Data availability.} The package contains the extraction script, the
per-cell and per-item counts it produced from commit 73475012 of the authors'
repository, the analysis and figure scripts, the number-check script that
recomputes every figure in this note from the analysis output, and the
reference-resolution log. No text from the source study is included.

\textbf{Competing interests.} The author runs Rankfor.AI, a company that
measures how AI engines describe brands and sells that measurement. Whether
language models share one taste for text or each prefer their own bears on
that business, and the reader should weigh the analysis with that in mind. The
data are the original authors' and every number here recomputes from their
public counts. No funding was received for this work.

\textbf{Use of AI tools.} The author used an AI assistant (Claude, Anthropic) for
grammar and language editing, and to draw the figures from the analysis output
by script. No data in this note were labelled or generated by AI tools. The
author reviewed all of this output and takes full responsibility for the content.

\bibliography{references}

\end{document}